# The Origin of Life from Primordial Planets

Carl H. Gibson [1,2]

[1] University of California San Diego, La Jolla, CA 92093-0411, USA
[2] cgibson@ucsd.edu, http://sdcc3.ucsd.edu/~ir118

Rudolph E. Schild [3,4]

[3] Center for Astrophysics, 60 Garden Street, Cambridge, MA 02138, USA
[4] rschild@cfa.harvard.edu

N. Chandra Wickramasinghe

Cardiff Centre for Astrobiology, Cardiff University, 2 North Road, Cardiff CF10 3DY, UK
NCWick@googlemail.com

**Abstract:** The origin of life and the origin of the universe are among the most important problems of science and they might be inextricably linked. Hydro-gravitational-dynamics (HGD) cosmology predicts hydrogen-helium gas planets in clumps as the dark matter of galaxies, with millions of planets per star. This unexpected prediction is supported by quasar microlensing of a galaxy and a flood of new data from space telescopes. Supernovae from stellar over-accretion of planets produce the chemicals (C, N, O, P etc.) and abundant liquid water domains required for first life and the means for wide scattering of life prototypes. The first life likely occurred promptly following the plasma to gas transition 300,000 years after the big bang while the planets were still warm, and interchanges of material between planets constituted essentially a cosmological primordial soup. Images from optical, radio, and infrared space telescopes suggest life on Earth was neither first nor inevitable.

## 1. Introduction

After a long and careful study of cosmic panspermia with one of us (NCW), Fred Hoyle suggested in 1980 that «...the cosmic quality of microbiology will seem as obvious to future generations as the Sun being the centre of the solar system seems obvious to the present generation…". Panspermia is the ancient idea that life seeds are distributed everywhere in the cosmos (Anaxoragas, 500 BC). Panspermia theories (Hoyle and Wickramasinghe 1977, 1982, 2000) and the field of astrobiology have been greatly hampered by their lack of feasible mechanisms for the large-scale transmission of microbiological information between planets within the standard (ΛCDMHC) cosmological model where planets are produced by stars and stars are produced by gas. The collision time between planets of different stars of a galaxy by the cold-dark-matter (CDM) hierarchical clustering (HC) model with dark energy (Λ) is roughly a billion times the age of the present universe. Because a key result of HGD is that the dark matter of galaxies is dense clumps of ancient planets, each of which is a potential host for life, this first order argument against astrobiology and panspermia (Wickramasinghe 2010, Wickramasinghe & Napier 2008, Wickramasinghe et al. 2004) is removed.

By ΛCDMHC, (the standard cosmological model) it is highly improbable that life could be widely transferred in the cosmos, and impossible for life to begin. The extreme complexity of the simplest living microorganism suggests spontaneous creation (abiogenesis) is impossible without templates. Hydro-gravitational-dynamics (Gibson 1996, Nieuwenhiuzen et al. 2009) provides the fluid mechanical mechanisms and the large time periods that ΛCDMHC lacks for life on Earth to exist in its present state. Galaxy microlensing of a quasar discussed by one of us (RES) reveals planetary mass twinkling frequencies that support the conclusion that planets constitute the missing mass of galaxies (Schild 1996). HGD-cosmology predicts the optimum time for first life to appear is soon after plasma became gas 13.7 Gyr ago (Gibson & Wickramasinghe 2010). HGD suggests that early seeds of life should be rapidly and widely scattered on cosmic scales and collected gravitationally by the trillion planets expected within each clump. Evaporated atmospheres of the frozen gas planets are photo-ionized to detectably large scales in planetary nebula such as Helix. The Oort cloud of long period comets is revealed to actually be proto-comet-planets at the boundary of a cavity in the proto-globular-star-cluster (PGC) clump of dark matter planets of size reflecting the large primordial mass density of the planets and the mass of the star or binary star formed in the center that have converted the H-He comets to a rapidly spinning carbon white dwarf.

Evidence from new telescopes, and improved old telescopes, show the standard cosmological model ΛCDMHC requires major modification to include effects of modern fluid mechanics (Gibson 2009ab). Including viscosity, diffusivity and stratified turbulent transport processes requires a new cosmology started by big bang turbulence and its fossils. From HGD, cold-dark-matter CDM does not exist and neither does the standard (non-turbulent, inviscid) dark energy (Λ) invented by Einstein to produce a static universe by antigravity. From HGD, the dark matter of galaxies is primordial planets in clumps that dominate the mass of the interstellar medium and the formation, evolution and death of stars. The mass of non-baryonic dark matter appears to be ~30 times more in the form of primordial neutrinos, weakly collisional and very hot, diffused away from $10^{22}$ m galaxy halo scales to $10^{23}$ m galaxy cluster halo scales and larger. Such matter is dynamically irrelevant to galaxy and star evolution.

According to our favoured, and in our view correct cosmological model, HGD, planets form all the stars as well as their comets in a sequence of powerful binary mergers that continually recycle life seeds, fertilizer and ecosystems. This also, incidentally, explains why most stars are binary. Radio telescope signals from these $10^5$-times-a-day events have recently been detected (Ofek et al. 2009) and identified as either an inexplicable galactic population of neutron stars or near-by exotic interstellar explosions without optical or infrared counterparts. The gravitational explosive power of a ten-Jupiter merger occurring in a three-month period corresponds to that of a million suns, and can account for the radio telescope signals observed, and also produce the recycling of materials of the larger and smaller objects produced. Together the planets, moons and comets of a PGC collect and redistribute life-giving stardust and life prototypes throughout the galaxy and galaxy clusters.

Runaway accretion of planet-proto-comets leads to supernova deaths of stars that expel thousands of planets throughout the primordial planet clumps along with stardust fertilizer, water, carbon and life infections and organic waste products of other planets. According to HGD, life is common in the cosmos and has existed throughout all space at galaxy and galaxy cluster scales since soon after the gas epoch began. The ability of living organisms to efficiently convert most of the carbon of a planet to complex organic compounds that absorb and reradiate light differently than abiological carbon compounds may explain the ultra-luminous infrared galaxy (ULIG) phenomenon, for example. A new class of blue galaxies detected to redshift z >10 appears to be dust free, possibly because life and its wastes had not yet had time to develop.

The plan of the paper is to first discuss the formation of clumps of primordial planets at the plasma to gas transition in the very early universe, since this is where the first life forms probably appeared. The following section compares non-standard (correct) HGD and the standard LCDMHC cosmologies in relation to mechanisms for the origin of life and panspermia. Finally, discussion and conclusion sections are presented.

## 2. The formation of primordial planets in clumps

Observations support a big bang event (Peebles et al. 2009), where spectral lines of distant objects are red shifted by a stretching of space in an expanding universe according to Einstein's equations of general relativity (Peacock 2000). General agreement exists (Weinberg 2008) that the big bang event was hot and produced a primordial plasma of protons, alpha particles and electrons with mass-energy dominated by mass at about $10^{11}$ seconds (3,000 years) after the big bang event and the plasma to gas transition at $10^{13}$ s (300,000 years). Unfortunately, strong fluid mechanical simplifications necessary in the early days of cosmology have been continued to this day. It has been assumed that the fluids of the early universe are frictionless, irrotational, linear and ideal. They are not. The relevance of fluid mechanics to astrobiology arises from the HGD prediction that the dark matter of galaxies is frozen H-He planets in clumps with appropriate admixtures of biogenic elements (Gibson & Wickramasinghe 2010). How does this happen?

Viscosity is neglected in the standard cosmology model based on the Jeans 1902 acoustical criterion for gravitational structure formation in a gas (Jeans 1902). By this criterion, gravitational structure formation begins when the Jeans length scale $L_J = V_S/(\rho G)^{1/2}$ becomes smaller than the horizon scale, or scale of causal connection, $L_H = ct$, where $V_S$ is the speed of sound, $\rho$ is the density, G is Newton's gravitational constant, c is the speed of light, and t is the time. Jeans used the inviscid Euler equations, linear perturbation theory, and neglected diffusivity effects that turn out to be crucial. These strong assumptions reduce the momentum conservation equations to a linear acoustic form and the Jeans criterion for structure formation. No gravitational structure can form during the plasma epoch $10^{11}$ s $< t < 10^{13}$ s because the Jeans length scale $L_J = V_S/(\rho G)^{1/2}$ is less than the scale of causal connection $L_H = ct$ throughout this time period, where c is the speed of light and t is the time. The reason this is true is that the relevant density is that of the baryons that retain density fluctuations from big bang turbulence. The total density of the universe is dominated by weakly collisional non-baryons (neutrinos) that are so strongly diffusive that all nonbaryonic density fluctuations are smoothed away. The baryonic gravitational free fall time $(\rho G)^{-1/2}$ is ~2.7 times larger than the critical value, causing failure of the Jeans criterion and the invention of cold dark matter.

Viscous forces prevent structure formation in the plasma epoch while $L_H < L_{SV} = (\gamma \nu/\rho G)^{1/2}$, where $\gamma$ is the rate of strain, $\nu$ is the kinematic viscosity and $L_{SV}$ is the Schwarz viscous-gravitational length scale (Gibson 1996). The kinematic viscosity $\nu$ of the plasma is enormous (Gibson 2000), permitting proto-supercluster-void formation to begin when $L_H = L_{SV}$ at $t = 10^{12}$ s. Photon viscosity ceases when the cooling plasma turns to gas at $t = 10^{13}$ s. Because the gas viscosity is $10^{13}$ smaller, the fragmentation mass decreases by a factor of $(10^{13})^{3/2}$; that is, from galaxy mass $10^{43}$ kg to earth mass ~$10^{24}$ kg. An additional gravitational instability occurs at $L_J$ due to a mismatch between heat transfer at the speed of light and pressure transfer at the speed of sound, so fragmentation at the plasma-gas phase change produced $10^{36}$ kg Jeans-mass clumps of earth-mass planets. These are proto-globular-star-cluster (PGC) clumps of primordial-fog-particle (PFP) planets (Gibson 1996).

The PGC density is the baryonic density at the $10^{12}$ s time of first structure $\rho_0 = 4 \times 10^{-17}$ kg m$^{-3}$ that is observed in old globular star clusters (OGCs) of all galaxies. Stars form within such PGC clumps by mergers of the planets. If stars do not form but the planets freeze as the expanding universe cools, these clumps of frozen planets become the dark matter of galaxies and most of the mass of the interstellar medium. Objects of mass M leave an $(M/\rho_0)^{1/3}$ size empty Oort cavity of missing planets in a PGC clump, which grows as the central object grows, to a $3\times10^{15}$ m size for one star. Proto-comets from the cavity boundary continuously feed the growth of the inner star and will eventually cause its death when its mass exceeds the Chandesekhar limit of 1.44 solar mass when the star is a white dwarf. As evidence, consider the Spitzer space telescope infrared image of the Helix planetary nebula shown in Figure 1. The central star of Helix is a rapidly spinning white dwarf that converts the accreted planet «proto-comets» to gas. Part of the gas is converted by the star to more carbon, and part is emitted at the poles as a plasma jet that evaporates any planets it encounters. The photoionized atmospheres and their wakes are large enough to be visible, between $10^{13}$ m and $10^{14}$ m in diameter (an astronomical unit is $\sim10^{11}$ m, the size of the solar system).

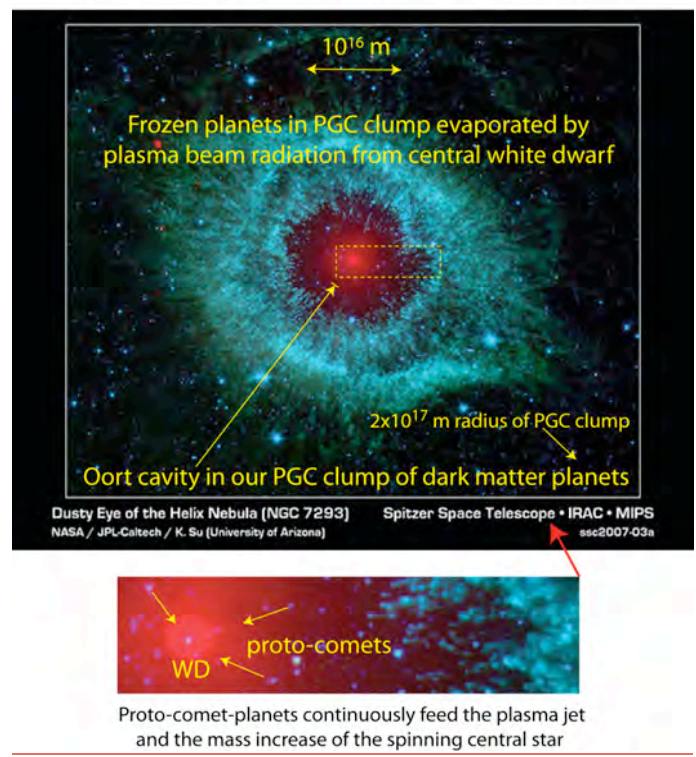

Fig. 1. Spitzer Space Telescope infrared view of Helix PNe at $6.7 \times 10^{18}$ m. The central white dwarf converts the hydrogen-helium frozen gas of proto-comets shown in the dashed box (bottom insert) to carbon of the shrinking star, and to a plasma jet that partially evaporates the ambient dark matter planets of the PGC clump of dark matter planets. The red dust cloud about the white dwarf is presumably left by the accreted proto-comets as they evaporate near the star, feeding its growth toward instability.

All of the planets shown in Fig. 1, as well as the proto-comets and moons of planets forming orbital systems deep within the Oort cavity, are potential hosts for life.

Figures 2 shows optical and infrared evidence of primordial planets and their (presumably) biologically generated dust in PNe Helix (Meaburn & Boumis 2010, Matsuura et al. 2009). The differences in opacity between radio, infrared, and optical frequencies offers a powerful tool for explor-ing the probabilities of life formation. Evidence from new infrared space telescopes Spitzer, Planck and Herschel shows that optically opaque dust is not always formed where planets exist on proto-globularstarcluster (PGC) scales. If dust clouds containing such polycyclic aromatic hydrocarbon (PAH: oils) and anthracite (coal) spectral signatures have a biological origin, then abiological processes must be extremely improbable on average planets like Earth, considering the variety of panspermic mechanisms and long time periods offered by HGD cosmology.

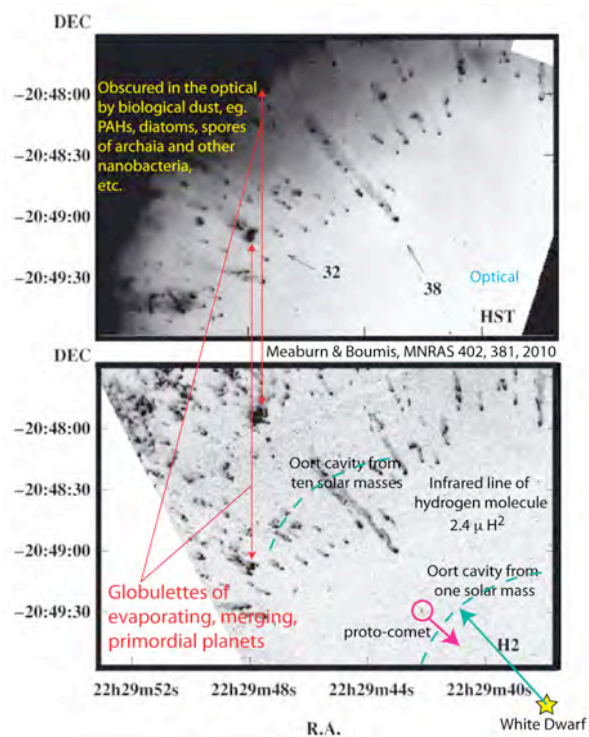

Fig. 2. Images from the Hubble Space Telescope of the Helix planetary nebula in the optical (top) and 2.4 μ $H^2$ infrared (bottom), from Meaburn & Bourmis 2010. Dust from evaporating frozen gas planets at the Oort cavity boundary (bottom) obscures clumps of merging planets (globulettes, Gahm et al. 2007) in the optical (top). Most hydrogen wakes point toward the central white dwarf star (bottom right). The planet numbered 38 (top) has a 2.4 x $10^{14}$ m diameter atmosphere with mass over $10^{26}$ kg, suggesting a massive (~Jupiter) evaporating frozen planet within.

Figure 2 (top) is an optical image (Meaburn & Boumis 2010) of planetary nebula Helix from the Hubble space telescope, showing evaporating planets at the Oort cavity boundary surrounding the central white dwarf compared to 2.4 μm $H^2$ line images unobscured by dust (Matsuura et al. 2009).

Figure 3 (top) is the unobscured infrared Subaru telescope image (Matsuura et al. 2009) showing proto-cometary-planets, detectable from their relatively small $10^{12}$-$10^{13}$ m $H^2$ atmospheres, on their way to feed the central star, but invisible (bottom) in the dust-obscured HST optical image. Larger diameter $10^{13}$-$10^{14}$ m radiation heated planet atmospheres appear in both Fig. 3 images beyond the edge of the Oort cavity. Correspondingly larger radiation pressures accelerate such planets radially outward, expelling the more dusty planets surrounding the central star into the interior of the PGC planet clump along with whatever biological templates, organisms and wastes the planets contain.

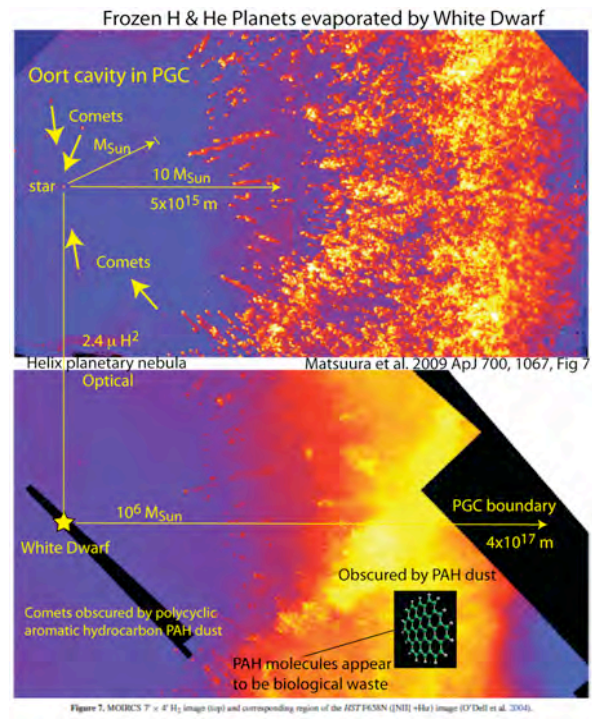

Fig. 3.  Images from the Subaru telescope of Helix planets and comets in the 2.4 μ $H^2$ frequency band (top) compared to the Hubble space telescope optical image (bottom) in at optical frequencies (Matsuura et al. 2009, fig 7).  Proto-comet-planets (top) fall toward the central white dwarf because their biological dust content, atmosphere size and radial radiation pressures are smaller.

Sizes of sphere radii containing one, ten and a million solar masses from the expression $(M/\rho_0)^{1/3}$ are shown in Fig. 3.  A proto-type PAH molecule is shown at the bottom of Fig. 3.  The structure is characteristic of a carbohydrate oil used for food by terrestrial organisms, and would be a biological waste product expelled by PNe radiation-evaporating biologically inhabited planets.

In the following we first compare the HGD and ΛCDMHC cosmological models emphasizing aspects that affect the formation and transmission of life.  We then examine the observational evidence, and provide discussion and conclusions.

### 3.  Hydrogravitational dynamics (HGD) versus the standard cosmological model

The standard dark-energy cold-dark-matter hierarchical-clustering (ΛCDMHC) model of cosmology is modified by modern fluid mechanics to produce the hydrogravitational dynamics (HGD) cosmological scenario of Figure 4abcde.

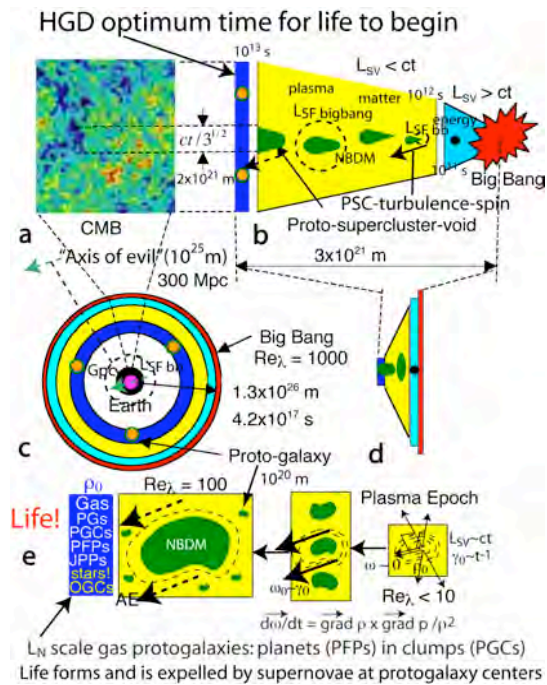

Fig.4. Early formation of gravitational structures supporting life according to hydro-gravitational-dynamics (HGD). At plasma to gas transition the entire baryonic plasma universe fragments at the gas Schwarz viscous scale $L_{SV}$ to form planets in clumps within protoclusters and protosuperclusters of protogalaxies.

The key difference is rejection of the Jeans 1902 sonic criterion for plasma gravitational structure formation in favor of photon viscosity (Gibson 1996, 2000, 2008). Cold dark matter (CDM) condensations are unnecessary and indeed physically impossible. The Jeans length scale $L_J = V_S t$ is larger than the scale $L_H$ of causal connection $ct$ throughout the plasma epoch, but the Schwarz viscous length scale $L_{SV}$ becomes smaller than $ct$ at a time of only 30,000 years ($10^{12}$ s), which is the time of first gravitational structure formation by fragmentation to form protosuperclustervoids and protosuperclusters (PSC).

Vorticity and weak turbulence at the expanding supervoid boundaries determines the morphology of plasma protogalaxies formed just before transition to gas (Gibson & Schild 2009). Fragmentation of protogalaxies occurs along plasma vortex lines at Nomura scales ($10^{20}$ meters) and morphology (Nomura & Post 1998) to form chain-galaxy-clusters (CGCs) clearly visible in the Hubble space telescope ultra deep field images (Elmegreen et al. 2005) and the Tadpole galaxy merger (Gibson & Schild 2002). The non-baryonic dark matter particles (neutrinos) shown in green diffuse to fill the expanding voids and expanding plasma (Nieuwenhuizen 2009). Fossils of the turbulent big bang epoch include spin at the strong force freeze out scale $L_{SF}$ stretched from meter scales at the end of inflation to $> ct$ until after the plasma epoch. The plasma density $\rho_0 = 4 \times 10^{-17}$ kg/m$^3$ at the time of first gravitational fragmentation $10^{12}$ s is preserved as the density of globular star clusters, and the initial fragment mass scale $\rho_0 (ct)^3$ of a thousand galaxies ($3 \times 10^{45}$ kg) is preserved as the baryonic mass of galaxy superclusters.

By the HGD scenario of Fig. 4abcde, the optimum time for the first life to form and be widely transmitted by cometary panspermia is $10^{13}$ seconds (300,000 years) after the big bang turbulence event when proto-planets in clumps first appear in protogalaxies. A turbulent big bang instability occurs at Planck scales, where gravitational, inertial-vortex and centrifugal forces produce the first turbulent combustion by extracting mass-energy from the vacuum by stretching turbulent vortex lines until the turbulence power is enhanced exponentially by gluon viscosity and the appearance of the first quarks (Gibson 2004, 2005). Vortex line stretching produces inflation to meter scales at Planck densities with an extraction of big bang mass-energy some forty orders of magnitude larger than the mass energy observable in our horizon $L_H = ct$, where $c$ is the speed of light and $t$ is the time since the big bang event. Because temperatures were so high, little entropy was produced except by the turbulence[1]. The resulting expansion and cooling of the big bang universe has a long time constant before the inevitable «big crunch» return of the extracted energy to the vacuum. Fossil turbulence[2] patterns of the big bang are frozen as the exponentially expanding universe carries structures to scales separated by distances exceeding $L_H$, also termed the scale of causal connection since the maximum speed of gravitational information transport is the speed of light. Because galaxy patterns observed in opposite directions are identical and outside each other's horizon suggests an inflationary expansion period followed the big bang.

Turbulence motions of the big bang have produced unambiguous fossil turbulence fingerprints in the temperature anisotropy patterns observed by cosmic microwave background space telescopes (Bershadskii 2006, Bershadskii & Sreenivasan 2002, 2003, Gibson 2010).

The density and rate-of-strain of the plasma at transition to gas at 300,000 years ($10^{13}$ s) are preserved as fossils of the time of first structure at 30,000 years ($10^{12}$ s), as shown in Fig. 4e. The plasma turbulence is weak at transition, so the Schwarz viscous scale $L_{SV} = (\gamma \nu / \rho G)^{1/2}$ and Schwarz turbulence scale $L_{ST} = \varepsilon^{1/2} / (\rho G)^{3/4}$ are nearly equal, where $\gamma$ is the rate-of-strain, $\nu$ is the kinematic viscosity, $\varepsilon$ is the viscous dissipation rate and $\rho$ is the density. Because the temperature, density, rate-of-strain, composition and thus kinematic viscosity of the primordial gas are all well known it is easy to compute the fragmentation masses to be that of protogalaxies composed almost entirely of $10^{24} - 10^{25}$ kg planets in million-solar-mass $10^{36}$ kg (PGC) clumps (Gibson 1996). The non-baryonic-dark-matter (NBDM, neutrinos) diffuse to diffusive Schwarz scales $L_{SD} = (D^2 / \rho G)^{1/4}$ much larger than $L_N$ Nomura scale protogalaxies, where $D$ is the large NBDM diffusivity with D ~ ν. The rogue-planets prediction of HGD was independently predicted (Schild 1996) by interpretation of a long series of careful quasar microlensing observations that have since been refined and confirmed, and extended, to show the planet mass objects must be in dense clumps.

---

[1] Turbulence (Gibson 1991, 2006) is defined as an eddy-like state of fluid motion where the inertial-vortex forces of the eddies are larger than any other forces that tend to damp the eddies out. Turbulence by this definition always cascades from small scales to large, starting at the viscous-inertial-vortex Kolmogorov scale at a universal critical Reynolds number. The mechanism of this «inverse» cascade is merging of adjacent vortices with the same spin due to induced inertial vortex forces that force eddy mergers at all scales of the turbulence. Such eddy mergers account for the growth of turbulent boundary layers, jets and wakes. A myth of turbulence theory is that turbulence cascades from large scales to small. It never does, and could not be universally similar if it did. Irrotational flows cannot be turbulent by definition. In self-gravitational fluids such as stars and planets, turbulence is fossilized in the radial direction by buoyancy forces. Radial transport is dominated by fossil turbulence waves and secondary (zombie) turbulence and zombie turbulence waves in a beamed, radial, hydrodynamic-maser action.

The optimum time of life formation in our favoured HGD cosmology is soon after the plasma to gas transition when the first stars begin to form by binary accretion of primordial planets. Life-forming chemicals will be first formed near protogalaxy centers where strong tidal forces cause turbulence and supernovas of large short-lived stars at turbulent Schwarz scales $L_{ST} = \varepsilon^{1/2} / (\rho G)^{3/4}$. At this time most of the baryonic matter in the universe would be in the form of heavy-element enriched primordial planets, and the typical separations between planets would be of the order of several AU. With radioactive nuclides from supernovae (eg. 26Al) causing melting in their interiors, and with frequent exchanges of material taking place between planets, the entire universe would essentially constitute a connected primordial soup. Life would have an incomparably better chance to originate in such a cosmological setting than at any later time in the history of the universe. Once a cosmological origin of life is achieved in the framework of our HGD cosmology, exponential self-repication and propagation continues, seeded by planets and comets expelled to close-by protogalaxies. Figure 5 contrasts biological processes likely in HGD versus ΛCDMHC cosmologies. The timelines for major events shown in Fig. 5 are very different, so the flood of new information coming from new space and ground based telescopes is rapidly falsifying ΛCDMHC. Both cosmologies assume a big bang initial event. Only irrotational, ideal fluid, frictionless, and adiabatic quantum fluctuations are assumed by the standard model, but HGD requires specific turbulence and fossil turbulence patterns with spin, and is supported by observations (Gibson 2009ab).

---

[2] Fossil turbulence (Gibson 1981, 1986, 1991, 1999) is a perturbation in any hydrophysical field caused by turbulence that persists after the fluid is no longer turbulent at the scale of the perturbation.

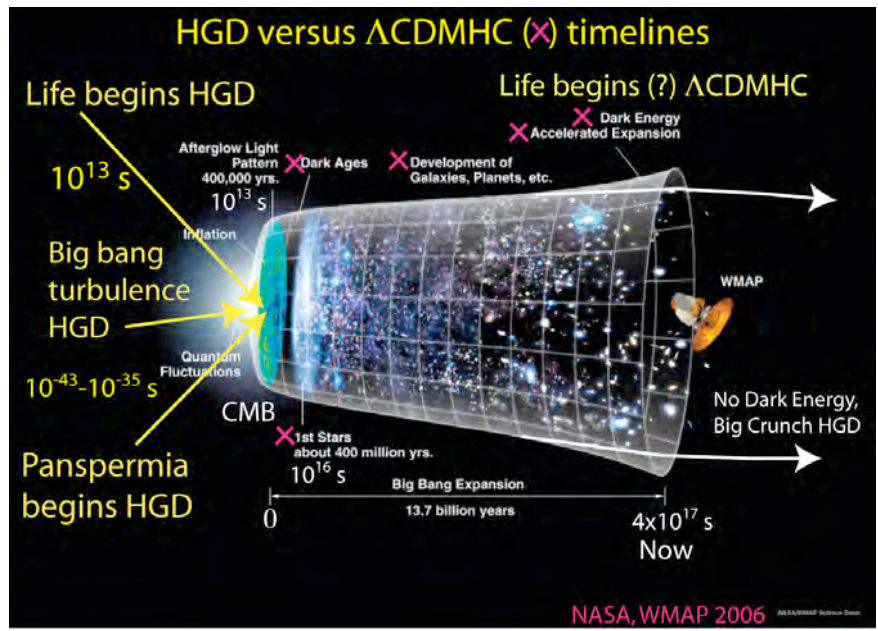

Fig. 5. Contrast between HGD and ΛCDMHC timelines for gravitational structures, and for life formation and panspermia. If abiogenesis can work in a time period of a few million years, then all the necessary chemicals, comfortable environments and means of dispersal of the resulting life templates are provided by HGD soon after the plasma to gas transition at $10^{13}$ seconds. Harsh early conditions of the standard model and its lack of primordial planets make life formation rare and more difficult to disperse.

As shown in Fig. 5, HGD conditions for life formation and transmission are optimum immediately after the first stars form and the first stars explode, soon after the plasma to gas transition. Because the first stars of ΛCDMHC are superstars that reionize the gas to plasma, and because planets and comets are produced in small numbers only after billions of years, life formation and intergalactic transmission of life templates by panspermia become highly unlikely. Big bang turbulent friction produces entropy and therefore a big crunch, rather than the open universe expected if one assumes cosmological constant dark energy (Λ), with big-bang-turbulence and gluon-viscous dark energies ($Λ_{HGD}$) as the necessary antigravity forces of the big bang.

Direct proof that planets in clumps are the source of all stars and the dark matter of galaxies is difficult. The nearest primordial planet to a star is about $5 \times 10^{15}$ meters, the distance to the Oort "cloud" of comets, representing the inner surface of the Oort cavity within a PGC clump produced when merging planets form a central star. Since the primordial planets are the mass of the Earth with the density of frozen hydrogen, the angular size of the planet from Earth is only ~$2 \times 10^{-10}$ radians. This is ~0.1% the resolution of the Hubble space telescope. Fortunately, when objects like dying stars begin to spin up and radiate, the dark matter planets come out of the dark because radiation can produce large gaseous atmospheres. The nearest planetary nebula to the earth is the Helix (Fig1. 1-3), where thousands of planetary objects have been observed by optical and infrared telescopes because their atmospheres have expanded to sizes up to ~$10^{13}$ meters. Radiation is caused by planet accretion and the rapid spinning of the white dwarf, pumping a powerful plasma jet that partially evaporates frozen gas planets at the inner boundary of the Oort cavity (Fig. 11, p 114, Gib-

son 2009b).

Because dark matter planets are much colder than stars, they are easier to see in the infrared frequency band than in the optical. Only ~5000 planets can be seen in Helix (Fig. 1-3) at optical frequencies by the Hubble Space Telescope, but ~40,000 were detected in the infrared by the Spitzer telescope (Fig. 1).

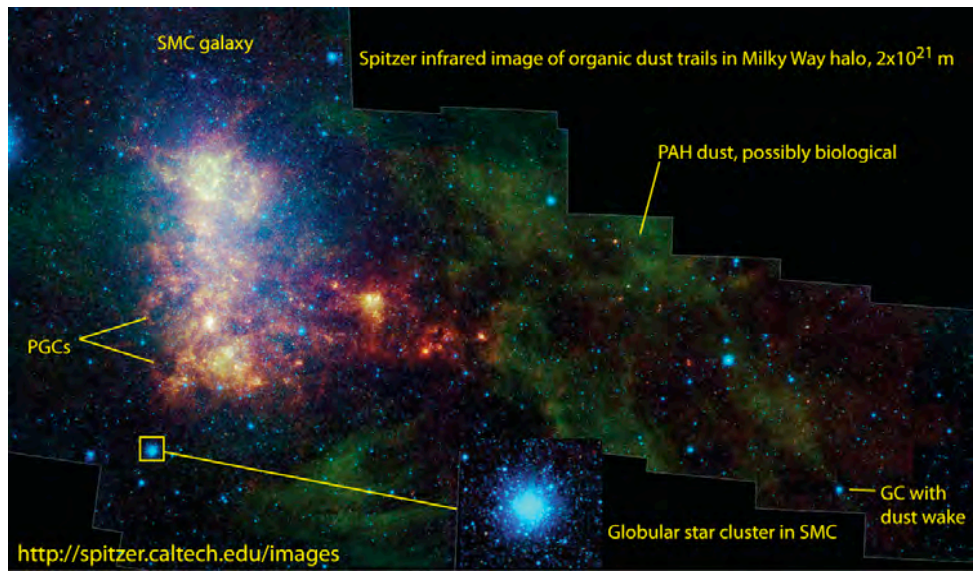

Fig. 6. Spitzer infrared telescope image of the Small Magellanic Cloud (SMC) galaxy at a distance of 70 kpc (~$2 \times 10^{21}$ m). Dark matter protoglobularstarclusters (PGCs) with fragmentation scale $(M/\rho_0)^{1/3} \sim 4 \times 10^{17}$ m show PAH dust wake evidence of life occurs only occasionally at PGC scales, and sometimes not at all, as seen for the SMC globular star cluster (bottom insert).

Figure 6 is an infrared Spitzer image of the Small Magellanic cloud. Possibly biological PAH dust wakes (green) appear behind some of the numerous globular star clusters (GCs), but not all. Dark PGCs with PAH dust appear (at left) in the SMC from optical backlighting with their primordial fragmentation scale $4 \times 10^{17}$ m, matching the globular star cluster size (center insert). The SMC clump of PGCs is well within the $10^{22}$ m diameter PGC dark matter halo of the Milky Way expected from HGD (and as observed, see Fig. 13).

## 4. Theory and observations

Important concepts missing in discussions leading to the standard cosmological model (Peacock 2000) are the absolute instabilities of turbulence, gravity and living processes. Starting from an initial condition at rest with small density fluctuations, structures will form in a gravitational free fall time $\tau_g = (\rho G)^{-1/2}$, where G is Newton's gravitational constant. Within a horizon length $ct$, mass will immediately start to move away from density minima and toward density maxima with exponentially increasing acceleration and viscous dissipation as the time approaches $\tau_g$. In an expanding universe the formation of voids is favored, as shown in Fig. 4bde.

The baryonic plasma density falls to near zero within a protosuperclustervoid within $10^{12}$ s and the void expands driven by gravity as a rarefaction wave limited by the sound speed $V_S=c/3^{1/2}$ in the plasma until the plasma to gas phase change at $10^{13}$ s. This is the HGD interpretation of the primary sonic peak with angular size about one degree observed by WMAP and other microwave background telescopes. The $0.4^o$ and $0.2^o$ sonic peaks are interpreted as wrapping, by fossil big bang turbulence spin, of secondary turbulence vortex lines, around the boundaries of the expanding protosuperclustervoids (Gibson 2010).

A preferred spin direction has been identified on the sky, toward which x-ray galaxy clusters are moving and toward which point the spin axes of the Milky Way and our local galaxies. It is also the direction of all low wavenumber spherical harmonic wave vectors of the CMB (Schild and Gibson 2008). It explains large extragalactic magnetic fields detected at superclustervoid scales ($10^{25}$ m, 300 Mpc) by the Fermi γ-ray space telescope (Neronov & Vovk 2010) as Tev γ-ray interactions with B-fields in large rotating voids expected from HGD but not ΛCDMHC.

To describe big bang turbulent combustion and inflation requires discussion of some basic properties of turbulence and fossil turbulence at the beginning of time, Gibson (2004, 2005). Newtonian gravitation is modified by relativistic effects where pressures are of order $\rho c^2$ and may be enormously negative from the effects of the induced non-turbulent energy cascade of Planck gas and radial viscous and anti-gravity turbulent inertial vortex forces, as shown in Figure 7, particularly at the strong force freeze out phase change, where the Planck fireball cools from the Planck temperature of $10^{32}$ K to $10^{28}$ K so that quarks and gluons (that transmit strong forces) become possible.

The big bang turbulent combustion instability is extremely efficient and breaks symmetry for the universe created with the first prograde accretion of a Planck anti-particle and a spinning Planck particle-antiparticle pair. From the Kerr metric, 42% of the rest mass energy of each prograde accretion is converted from gravitational potential energy to the production of more Planck particles (Peacock 2000, p61, eq. 2.106), always retaining a slight statistical bias toward the production of matter rather than antimatter in the big bang remnants.

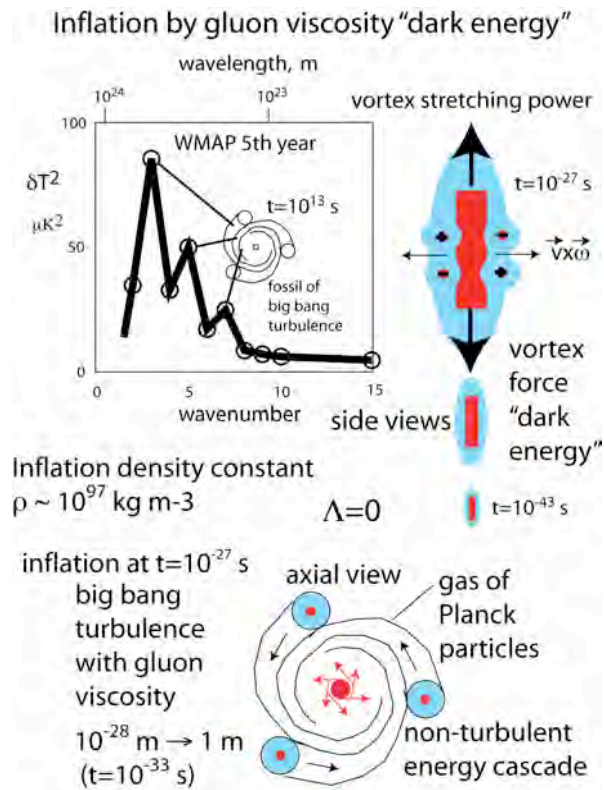

Fig. 7. Evidence for big bang turbulence driven inflation appears at the smallest wavenumber (largest scale) temperature anisotropies observed by the WMAP 5[th]. microwave background telescope team (top), interpreted here as a fossil of a non-turbulent kinetic energy cascade induced by big bang turbulent vortices (bottom).

During inflation, the expansion of space does work by stretching Planck-Kerr vortex lines and during inflation (Fig. 7 bottom) opposing gluon viscous stresses, exponentially creating yet more space-time and mass-energy. Secondary turbulent vortices wrap around the central vortex tube with alternating spin directions, creating powerful antigravity inertial-vortex-force and gluon-viscosity "dark energy". Gluons are the fundamental particles (bosons) that transmit the strong force which bind quark particles together in an atomic nucleus. Larger scale momentum transport from free gluons within the quark gluon plasma vastly increases the viscosity of the big bang fireball, converts total pressure from positive to negative, and powers exponential inflation of space and mass-energy (Fig. 7, bottom). The exponentially increasing negative power is balanced by exponentially increasing mass-energy during inflation, keeping constant mass-energy density near its Planck value of $10^{97}$ kg m$^{-3}$. Turbulent dissipation rates are high, but little entropy is produced at Planck temperatures $10^{32}$ K decreasing to $10^{28}$ K at strong force freeze out. Thus the turbulent big bang powered inflationary expansion of space is nearly adiabatic (flat) and will remain so for some time before its inevitable collapse.

For the combustible fluid of turbulent Planck particles and quark-gluon plasma of Fig. 7, the energy momentum tensor of Einstein's field equations is proportional to a frictional lost work terms (antigravity turbulent and gluon-viscosity "dark energies") rather than the inviscid, adiabatic term $\Lambda$, the Einstein cosmological constant. The source of anti-gravity is the turbulent inertial-vortex forces of adjacent vortex lines of opposite spin in the Planck gas surrounding the central vortex, stretched out in pairs with opposite spin, like the streamwise vortices of Langmuir cells in a wind driven boundary layer or the wingtip vortices of an airplane. Such forces cause smoke rings and wingtip vortices to translate and expand, and cause adjacent vortices with the same spin to merge. Stretching vortex lines in tornadoes and hurricanes accounts for the cascade of turbulent kinetic energy from small scales to large in these vortex structures, feeding on the induced non-turbulent kinetic energy cascade from large scales to small.

Figure 8 shows recent infrared HST evidence (right top) of vortex-line chain-clusters seen axially in Stephan's Quintet and in side view (bottom). See Fig. 13 below for the Tadpole image.

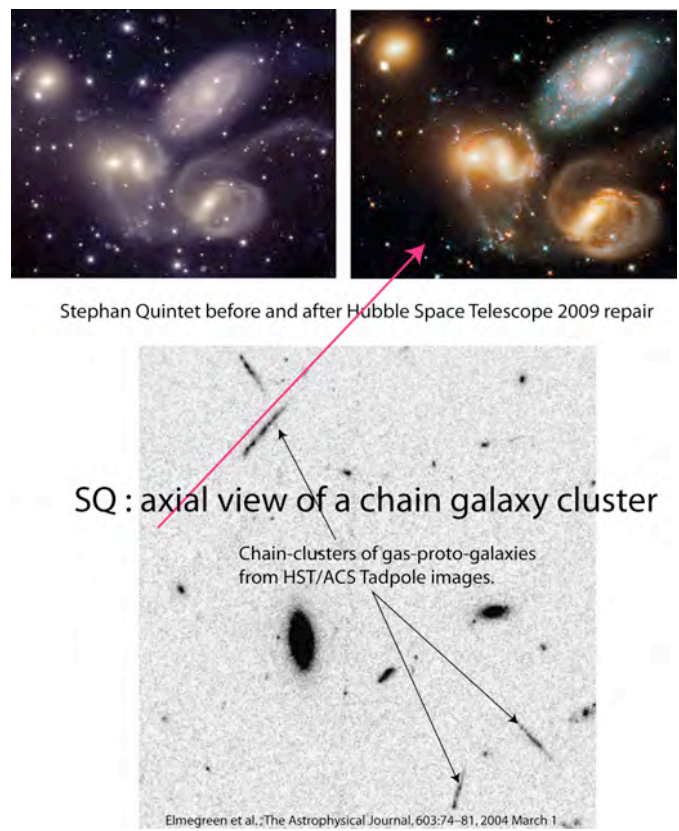

Fig. 8. Evidence that galaxies form as chain galaxy clusters CGC along plasma epoch turbulent vortex lines is provided by Stephan's Quintet. The repaired Hubble Space Telescope ACS3 and spectrometer (top) make the blue shift of the nearest galaxy obvious, as well as the interpretation (Gibson 2009ab) that SQ is a CGC axial view. Four CGC side views (bottom) appear in the HST image of the Tadpole galaxy merger.

The Herschel space telescope and the Planck space observatory were launched May 14, 2009, and are now providing high resolution infrared images from orbits at the second Lagrange point $1.5 \times 10^{12}$ m from Earth. Figure 9 shows one of the first Herschel images, from the Eagle star forming region of the Milky Way PGC accretion disk.

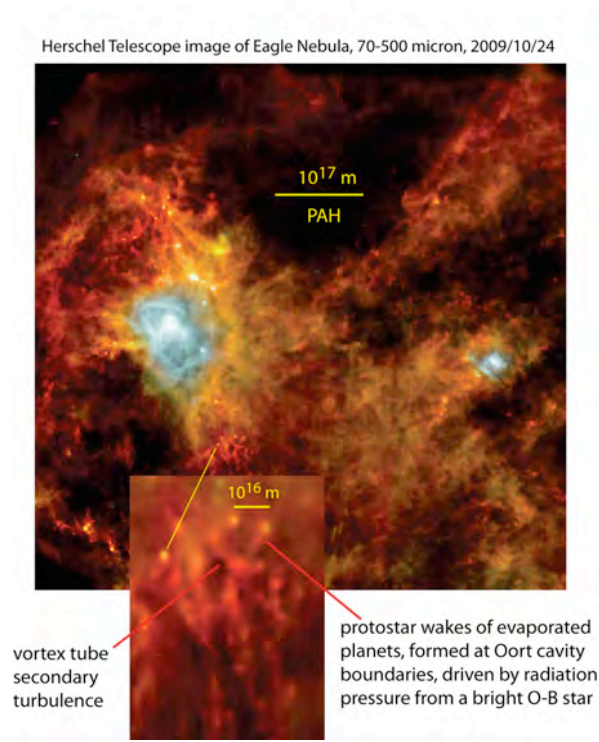

Fig.9. Herschel space telescope image from the Eagle star forming region (box Fig. 10). Bright infrared objects in the bottom insert are interpreted as large planets and brown dwarf protostars with warm earth-mass planet wakes evaporated at smaller Oort cavity boundaries than those shown in Fig. 1ab. The dark PAH obscured region at the top is interpreted as a PGC clump of strongly life-infected planets, reduced in size and distorted by the strong turbulence of this star forming region.

The Eagle star forming region is in the partially turbulent, stably stratified Milky Way accretion disk formed as frozen PGCs evaporated from the $10^{20}$ m Nomura scale protogalaxy core into the $10^{22}$ m diameter dark matter halo accrete on the disk plane. Figure 10 is a recent Planck space observatory image of what is described as a "tapestry of cold dust". The Herschel image of Fig. 9 is from the region shown by the dashed yellow box at lower left of Fig. 10, just above the disk.

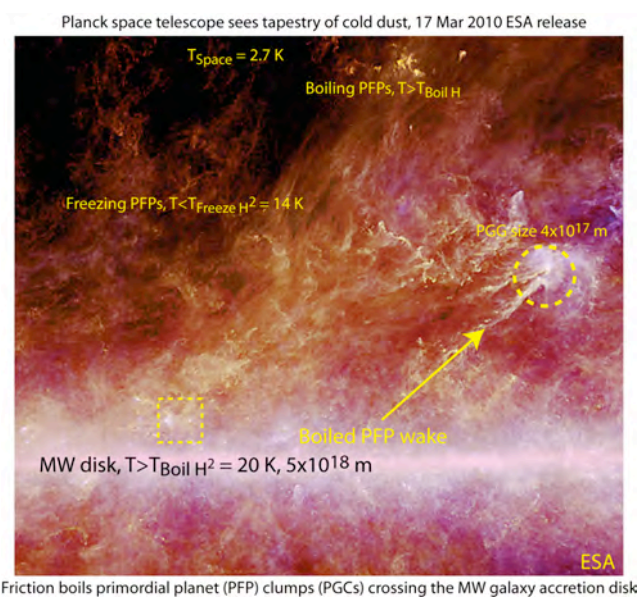

Fig. 10. Planck telescope ESA release image March 17, 1010. Measured temperatures reveal PGC wakes of merged planets triggered into formation by the PGC center of gravity as it moves through the Milky Way disk, as shown by the arrow. Fig. 6 from the dashed box star forming region shows evidence from PAH dust of turbulent mixing to smaller size of the PGCs.

Remarkably the dust temperatures of Fig. 10 are close to the 14-20 $^{\circ}$K freezing-boiling points of hydrogen, suggesting frictional forces of the accretion disk are responsible for evaporating the frozen planets of the PGC traversing the disk. Figure11 shows an infrared Herschel image of the optically dark "coalsack" dark cloud in the Crux southern hemisphere constellation (bottom insert).

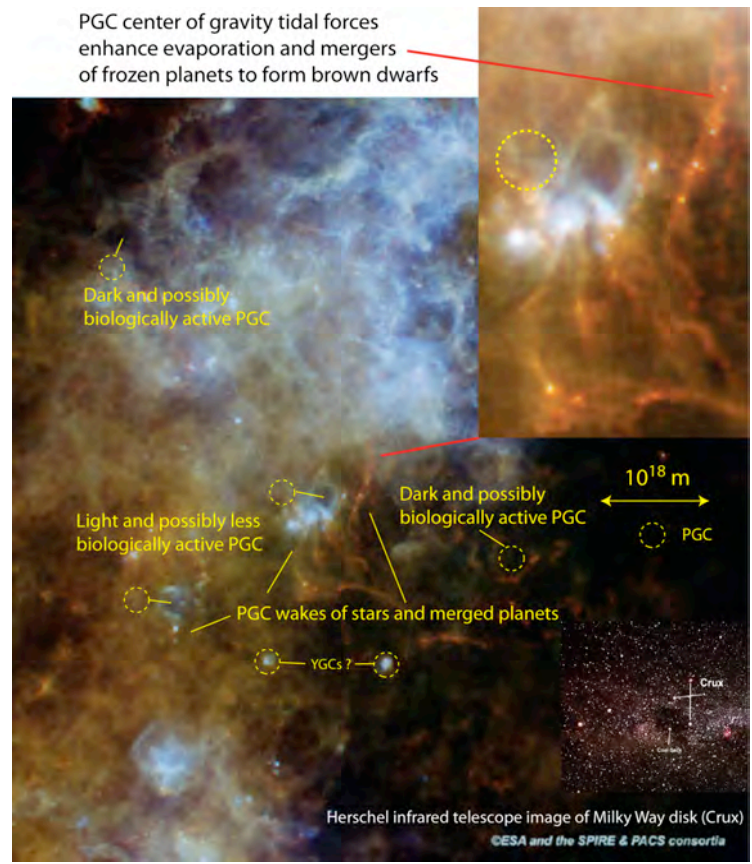

Fig. 11. Herschel space observatory infrared image of the optically dark "Coal Sack" in the Crux region of the Milky Way disk. Protoglobularstarcluster (PGC) scales are shown by dashed circles. Some are dark and some are light. Many are fringed by brightness and show wakes of bright clumps suggesting enhanced PFP planet mergers to form larger planets and gravitationally heated mini-brown-dwarf stars. Darkness and brightness of the PGC clumps is interpreted as PAH evidence of biological activity.

We consider these few infrared space telescope images to be only the first trickle in a future flood of information relevant to HGD predictions about the origin of life from primordial planets.

## 5. Discussion

An accumulation of evidence in a variety of frequency bands from a variety of very high resolution and highly sensitive modern telescopes supports the hypothesis that the dark matter of galaxies is primordial planets in proto-globular-star-cluster clumps, as predicted from HGD (Gibson 1996) and inferred from quasar microlensing (Schild 1996). All stars form from these planets so all star models and planetary nebulae models must be revised to take the effects of planets and their brightness and dimness into account. Figure 12 (Gibson 2010, Fig. 1) shows evidence of PGCs as the dark matter of the Milky Way Galaxy in a variety of frequency bands from microwave to infrared (Veneziana et al. 2010).  The DIRBE image permits a planet mass estimate exceeding that of stars.

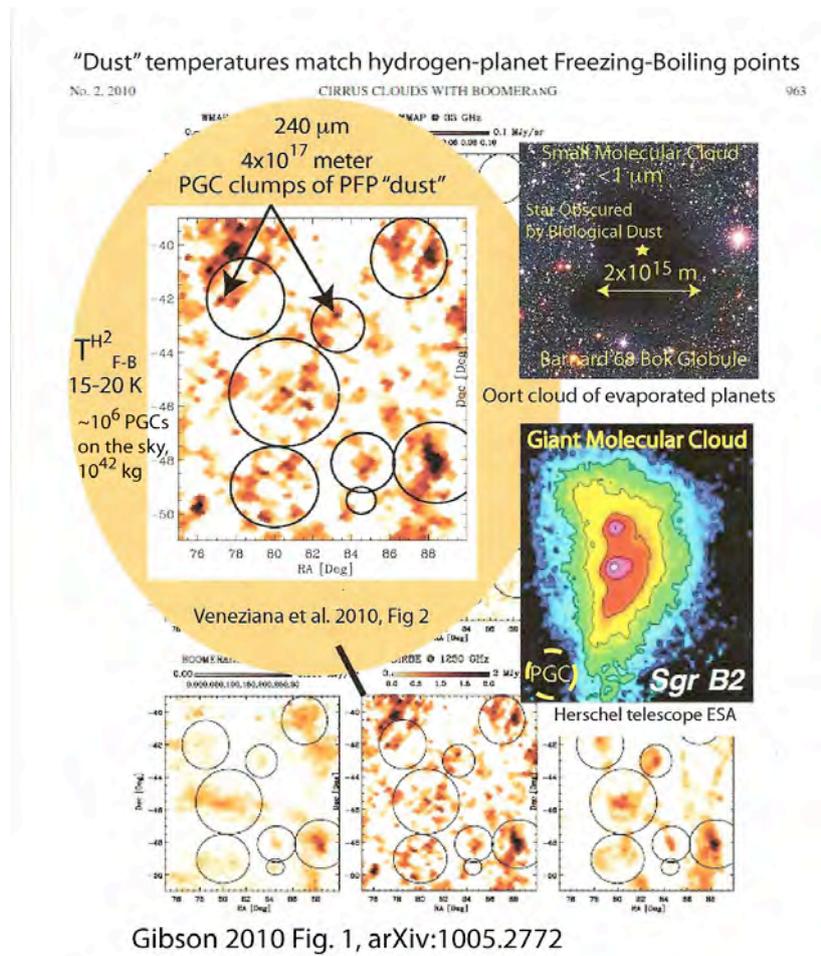

Fig. 12.  Microwave background images (Veneziana et al. 2010, figs. 1&2) from many platforms show redundant images of "cirrus dust clouds" with "dust" temperatures matching frozen hydrogen planet frozen-boiling temperatures $T_{FB}$ = 15-20 K.  The DIRBE 240 μm image (orange oval) indicates ~$10^6$ PGC "dust" clumps similar to the Sgr B2 GMC (right center).  A small molecular cloud in the visible suggests biological dust ejected from evaporated hydrogen-planets surrounding an Oort cavity (right top).

Fig. 12 leaves little room for doubt that the dark matter of the Milky Way Galaxy is PGC clumps of frozen hydrogen planets.  The "cirrus clouds of dust" (Veneziana et al. 2010 and other authors) are suspiciously warm in the range of triple point temperatures of frozen and boiling hydrogen expected if the "dust" of the cirrus clouds is planets in the Earth to Jupiter mass range, caused to change phase by their close proximity in planetary nebulae to stars formed by the planets within the planet clumps, and fed to dangerously large masses by a constant diet of comets now made visible by the new infrared space telescopes in Figs. 1-3.  A surprising discovery of a significant excess millimeter to submillimeter spectral emission of unknown nature from the large and small Magellanic clouds (Israel et al. 2010) implies massive (PGC) clumps of cold "dust" (primordial planets).

Figure 13 is the Tadpole galaxy merger (UGC 10214, VV29) Hubble Space Telescope archive high

resolution image, clearly showing merging galaxy fragments on a frictional track through the baryonic dark matter halo. Galaxy VV29a in the head of the Tadpole has the usual Nomura scale $L_N$ of old globular cluster (OGC) stars reflecting the size of its protogalaxy. As the PGC clumps of planets freeze the planets and clumps become increasingly collisionless and diffusive, forming the $L_{SD}$ scale baryonic dark matter (BDM) halo expanded by a factor of 80 in its radius. The non-baryonic-dark-matter halo (neutrinos) diffuses to a much larger radius. Even though the NBDM has ~30 times more mass than the BDM, its contribution to the central galaxy density is small.

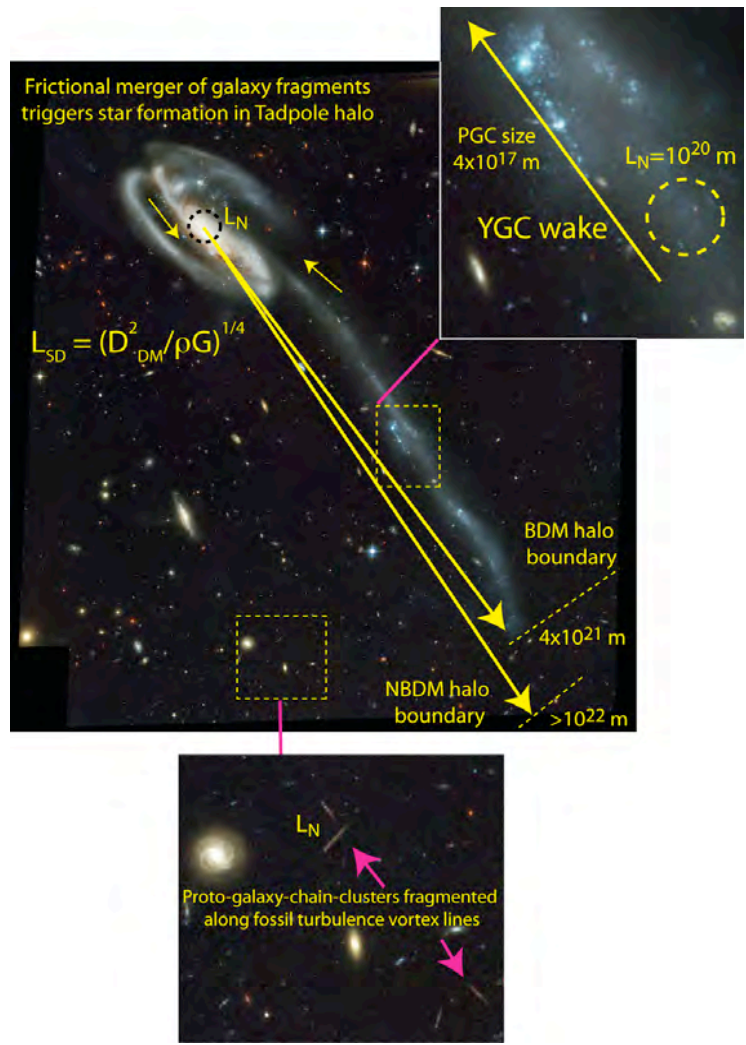

Fig. 13. Tadpole galaxy merger brings baryonic dark matter halo out of the dark. Nomura scale galaxy fragments trigger star formation (right insert) in young globular star clusters (YGCs). Tran et al (2003) identify 42 YGCs, precisely aligned along the star wake path toward frictional mergers of the fragments (Gibson 2008). Schwarz diffusive scales LSD are shown for the baryonic and nonbaryonic dark matter diffused from the $L_N$ scale protogalaxy to form BDM and NBDM halos. Background chain-galaxy-clustes (CGCs) are fragmented at $L_N$ scales along fossil turbulence vortex lines of the plasma epoch.

Turbulence produces post-turbulence-remnants (fossil turbulence) with structure in patterns that preserve evidence of previous events such as big bang turbulence and plasma epoch turbulence. Fossil-turbulence perturbations (Gibson 1981) guide the evolution of all subsequent gravitational structures. Numerous fatal flaws in the standard ΛCDMHC cosmology have appeared that can be traced to inappropriate and outdated fluid mechanical assumptions (Jeans 1902) that can be corrected by HGD (Gibson 1996, Schild 1996, Gibson & Schild 1999ab, 2002, 2007ab, Schild & Gibson 2008). Critically important advances in the understanding of fluid mechanics are the correct definitions of turbulence and fossil turbulence and the realization that all turbulence cascades from small scales to large (Gibson 1986, 1991, 2006).

Failure to understand these advances has also hampered fields of oceanography and atmospheric science, which have not yet recognized the importance of fossil turbulence, fossil turbulence waves, zombie turbulence and zombie turbulence waves as dominant physical mechanisms of terrestrial radial transports of hydrophysical fields, and for the preservation of information about previous turbulence (Keeler et al. 2005, Gibson et al. 2006, 2008). Enormous undersampling errors are characteristic of equatorial turbulence estimates in both the ocean and atmosphere from the same extreme lognormal intermittency factors that have prevented the MACHO, EROS and OGLE microlensing consortia from detecting primordial planets as the baryonic dark matter of galaxies (Gibson & Schild 1999ab). Figure 14 shows the effect of such undersampling errors on the EROS (Renault et al. 1997) MACHO (massive compact halo object) exclusion diagram.

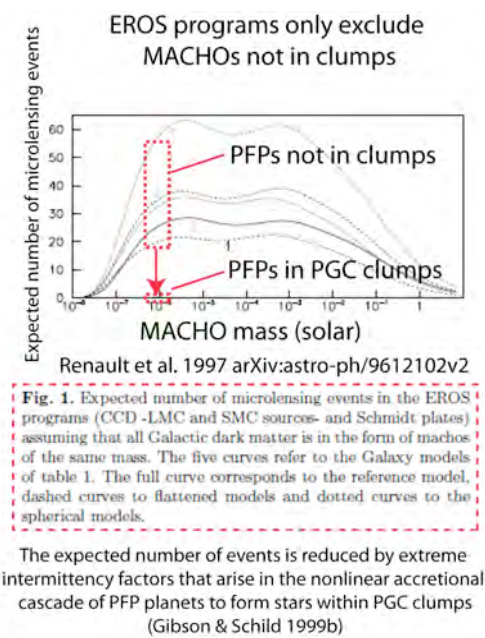

Fig. 14. Exclusion diagrams of the MACHO, EROS, and OGLE microlensing collaborations overestimate the number of events to be expected by assuming the MACHO objects (eg.: PFP planets) are not clumped.

As shown in Fig. 14, the EROS-1 collaboration expected 20-50 microlensing events from LMC and SMC star sources and saw none. However, all the Galaxy models of all the collaborations neglect the statistical effects of intermittent lognormal probability distributions (Gibson & Schild 199b).

We have seen that the enormous numbers of planets required by hydrogravitational dynamics have major biological consequences. Because planets in clumps are primordial components of dense and warm protogalaxies in protogalaxyclusters, it seems likely that the first life forms were also primordial. Because the efficiency of producing life from seeds is so much greater than producing life from inorganic chemical soups, it seems clear that the carbon based life forms observed on Earth reflect a cosmic ecosystem beginning soon after the plasma to gas transition, following the formation and death of the first stars. Old globular star clusters show the first stars were likely formed in a very gentle universe of warm gassy planets. The stars died when their overfed carbon cores exploded as SNe Ia events, giving the carbon, oxygen, calcium, nitrogen and other basic chemicals of life as we know it. Signatures of life on other planets are rapidly becoming abundant in the new infrared and radio telescope images. The amazing diversity and complexity of organic molecules found in the older-than-Earth Murchison meteorite (Schmitt-Koplin et al. 2010) and in the interstellar medium could in our model represent the disruption of biology and its waste products and nutrients from disintegrating frozen planets and comets.

In view of the grotesquely small improbability of the origin of the first template for life (Hoyle and Wickramasinghe, 1982) it is obvious that it would pay handsomely for abiogenesis to embrace the largest available cosmic setting. The requirement is for a connected set of cosmic domains where prebiology and steps towards a viable set of life-templates could take place and evolve. In the present HGD model of cosmology the optimal setting for this is in events that follow the plasma to gas transition 300,000 years after the big bang. A substantial fraction of the mass of the entire universe at this stage will be in the form of frozen planets, enriched in heavy elements, and with radioactive heat sources maintaining much of their interiors liquid for some million years. The close proximity between such objects (mean separations typically 10-30 AU) will permit exchanges of intermediate templates and co-evolution that ultimately leads to the emergence of a fully-fledged living system. No later stage in the evolution of the universe would provide so ideal a setting for the de novo origination of life. When re-frozen primordial planets are disrupted in the vicinity of stars within galaxies, panspermic dispersal of life occurs.

## 6. Conclusions

We conclude that standard models for the evolution of structure in the universe and standard models for the formation of life on Earth cannot be reconciled with the flood of new evidence from modern telescopes. The standard dark-energy-cold-dark-matter (ΛCDMHC) model is fatally flawed, doomed to failure by incorrect assumptions of collisionless, ideal, frictionless fluids and non-turbulent fluid motions. Modern concepts of turbulence, fossil turbulence, fossil turbulence waves and stratified turbulent transport result in vast differences between predictions for the formation of the universe, and the formation, evolution and morphology of gravitational structures with time. Hydrogravitational dynamics (HGD) and ΛCDMHC cosmologies are contrasted in Fig. 4. HGD leads immediately to conditions most favourable to the origin of life and its cosmic transport.

Primordial planet formation in clumps at the plasma to gas transition is the most important result of HGD cosmology with respect to the emergence of life on Earth. The planets merge to form stars, and pieces of planets form moons and comets as venues and vectors for life. Mergers of nearby primordial planets and their fragments are manifested by long duration radio transients lacking optical counterparts occurring with energies exceeding a Jansky $10^5$ times a year (Ofek et al. 2010) at distances $10^{18}$ to $10^{20}$ m, more than needed from simple binary mergers of planets to form Milky Way disk stars. Infrared and other telescopes (Spitzer, Planck, Herschel, HST) should be coordinated in future tests within the half hour to several day duration periods of such events to exploit this potential gold mine of information about the origin of stars and of life.

The abundance and degree of complexity of biologically relevant molecules in dense interstellar clouds are consistent with degradation of an all-pervasive cosmic biology that originated 300,000 yr after the big bang. Spitzer, Hubble and Subaru space telescope pictures of evaporating planets in Helix PNe in Fig. 1-3 show possible effects of biological waste on planet and comet heat and mass transport processes near a dying star. Dusty (biologically infected ) planets grow large atmospheres and are radially ejected by central white dwarf radiation pressure, while less dusty (less biologically infested) planets grow smaller atmospheres from reduced heating and fall into the star as proto-comets, proto-asteroids, and evaporated gas formed by friction, tidal interactions and radiation.

Milky Way disk images of infrared telescopes Spitzer and Herschel from which proto-globular-star cluster (PGC) clumps of planets with markedly different dust contents can be inferred. We hypothesize that the dust is biological in origin, and that PGC clumps without PCH (oil-like) dust to make them optically opaque are significantly less infected by carbon based organisms that rely on such carbohydrates for and polyaromatic hydrocarbons energy. Because frictional agitation of PGCs crossing and entrained by the Galaxy accretion disk is uniform, adequate chemical fertilizer and cosmic seeding to produce life in PGC planets in the disc of the galaxy should be available. The Planck space telescope more precisely confirms the boiling and freezing point temperatures of dark matter planets in dense clumps, with and without biological signatures, as the Galaxy dark matter. Summary plots in Figs. 12-14 overwhelmingly support the HGD claim that the dark matter missing mass of galaxies is primordial planets in primordial clumps.

It appears that hundreds of millions of years for a planet (like Earth) in galaxy disk conditions pro-vides no guarantee that life will occur from abiogenesis. If cosmologically generated primordial life is all-pervasive as we have argued, seeding should be a guaranteed route to life. It appears from recent studies of the Murchison meteorite (Schmitt-Kopplan, P. et al. 2010) that life forms on Earth have made only modest progress toward achieving the biochemical diversity displayed in the meteorite, even with continuous seeding. Without seeding, the observations suggest Earth would very likely not have formed life, our planet remaining sterile and lifeless to the present day.